\documentclass[aps,onecolumn,showpacs,tightenlines,superscriptaddress]{revtex4-2}
\usepackage{amsmath}
\usepackage{amsfonts}
\usepackage{graphicx}
\usepackage{epsfig}
\usepackage{color}
\usepackage{bm}
\usepackage[colorlinks,citecolor=blue]{hyperref}
\usepackage{diagbox}
\usepackage{multirow}
\usepackage{array}
\usepackage{makecell}

\begin{document}
\title{Feedback-Enhanced Driven-Dissipative Quantum Batteries\\ in Waveguide-QED Systems}
\author{Xian-Li Yin}
\affiliation{Department of Applied Mathematics, The Hong Kong Polytechnic University, Kowloon 999077, Hong Kong, China}
\author{Meixi Guo}
\affiliation{Department of Applied Mathematics, The Hong Kong Polytechnic University, Kowloon 999077, Hong Kong, China}
\author{Jian Huang}
\affiliation{Department of Physics, City University of Hong Kong, Kowloon, Hong Kong SAR}
\author{Heung-wing Joseph Lee}
\affiliation{Department of Applied Mathematics, The Hong Kong Polytechnic University, Kowloon 999077, Hong Kong, China}

\author{Guofeng Zhang}
\affiliation{Department of Applied Mathematics, The Hong Kong Polytechnic University, Kowloon 999077, Hong Kong, China}
\affiliation{The Hong Kong Polytechnic University Shenzhen Research Institute, Shenzhen, Guang Dong 518057, China}
\affiliation{Research Institute for Quantum Technology, The Hong Kong Polytechnic University, Hong Kong, China}

\begin{abstract}
Quantum batteries (QBs), acting as energy storage devices, have potential applications in future quantum science and technology. However, the QBs inevitably losses energy due to their interaction with environment. How to enhance the performance of the QBs in the open-system case remains an important challenge. Here we propose a scheme to realize the driven-dissipative QBs in atom-waveguide-QED systems and demonstrate significant improvements in both the stored energy and extractable work (ergotropy) of the QBs via feedback control. For a single-atom QB, we show that combining the measurement and coherent feedback controls enables nearly perfect stable charging under the weak coherent driving. For the QB array, the measurement-based feedback allows us to control different dynamical phases in the thermodynamic limit: (i) a continuous boundary time-crystal phase, where persistent periodic energy charge-discharge oscillations emerge despite the presence of the dissipation into the waveguide, and (ii) two stationary phases---one reaches full charge while the other maintains only small energy storage. This work broadens the scope of driven-dissipative QBs and provides practical strategies for enhancing their performance.
\end{abstract}

\date{\today}
\maketitle

\section{Introduction}
The rapid development of quantum technologies has created an urgent demand for efficient quantum energy storage devices~\cite{Sagawa15,Xuereb16,Auffeves22}. Quantum batteries (QBs), first proposed by Alicki and Fannes~\cite{Alicki13}, utilize quantum mechanical principles to achieve superior performance in charging power and work extraction compared to classical counterparts~\cite{Modi17,Polini19,Ghosh20,Santos20,Ferraro20,Nimmrichter21,Huangfu21,Rosa22,Yang22,Xue23}. Early theoretical studies in closed systems have revealed remarkable advantages through unitary charging protocols and collective quantum effects~\cite{Santos19,Dou20,Jing21,Rossini20,Kim22,Fusco16,Polini18,Crescente20,Dou24,Lu21,Liu21,Peng21}. However, the practical implementation of QBs remains hindered by the decoherence-induced energy loss and dissipative charging dynamics in open quantum environments. To address these challenges, various approaches have been proposed~\cite{Farina19,Barra19,Pirmoradian19,Chang21,Martins23,Horodecki23,Ahmadi24,Catalano24,Pokhrel25,Kamin20,Tabesh20,Wang22,Song24,Lu25,Quach20,Bai20,Shao21,Mitchison21,Shao22,Song25,Shao25}, such as utilization of non-Markovian effects~\cite{Kamin20,Tabesh20,Wang22,Song24,Lu25}, Floquet engineering~\cite{Bai20}, dark state~\cite{Quach20}, and feedback control~\cite{Shao21,Mitchison21,Shao22}. Experimental realizations of QBs have also been demonstrated across diverse platforms, including superconducting circuits~\cite{Zheng22,Hu22}, quantum dots~\cite{Buy23}, organic microcavities~\cite{Quach22}, and nuclear spins~\cite{Joshi22}.

Waveguide quantum electrodynamics (QED) systems~\cite{LiaoRV16,Roy17,Gu17,Sheremet21} have emerged as a promising platform for studying and realizing QBs. Existing implementations typically rely on passive energy transfer mechanisms, such as remote wireless charging mediated by atom-photon bound states~\cite{Song24,Lu25} or collective effect against the energy loss of the QBs~\cite{Tirone25}. These passive mechanism, however, fundamentally differs from the driven-dissipative regime~\cite{Farina19,Barra19,Pirmoradian19,Chang21,Martins23,Horodecki23,Ahmadi24,Catalano24,Pokhrel25}, where the external laser field continuously pumps the charger or QBs while the dissipation caused by the fields in the waveguide maintains. The competition between the driving and dissipation may create rich non-equilibrium dynamics~\cite{Fazio18,Carollo20,Prazeres21} that render conventional protocols ineffective. To address this, feedback control~\cite{Wiseman94,Wiseman09,Wilson15,Tufarelli13,Ciccarello19,Buonaiuto21} offers a pathway to not only mitigate the decoherence effects but also to actively exploit the driven-dissipative environment for enhancing the performance of QBs. The key open question we address is how to implement such control to boost the stored energy under continuous driving.

In this work, we propose a driven-dissipative two-level-atom QB scheme via coupling the atoms to a one-dimensional waveguide. Taking advantage of feedback control, the performance of both the single-atom QB and multiple-atom QB array can be significantly improved. For the charging process in the single-atom QB, it can be injected more stable energy when the measurement feedback (MFB) control~\cite{Wiseman94,Wiseman09,Buonaiuto21} is applied to the system. Furthermore, by incorporating the coherent feedback (CFB) control~\cite{Tufarelli13,Wilson15,Ciccarello19}, the performance of the QB is further improved even in the achiral-coupling case. In the collective charging process of the multiple-atom QB array, the synergistic effect between the MFB control and external driving gives rise to different dynamical phases in the thermodynamical limit. We find that the QB array can exhibit persisting oscillation in the continuous boundary time-crystal (BTC) phase~\cite{Fazio18} and store more stable energy in one of the stationary phases. Our results provide new insights into the design of QBs in open quantum systems.

%%%%%%%%%%%%%%%%%%%%%%%%%%%%%
\begin{figure}[t]
\centering
\includegraphics[width=0.4\textwidth]{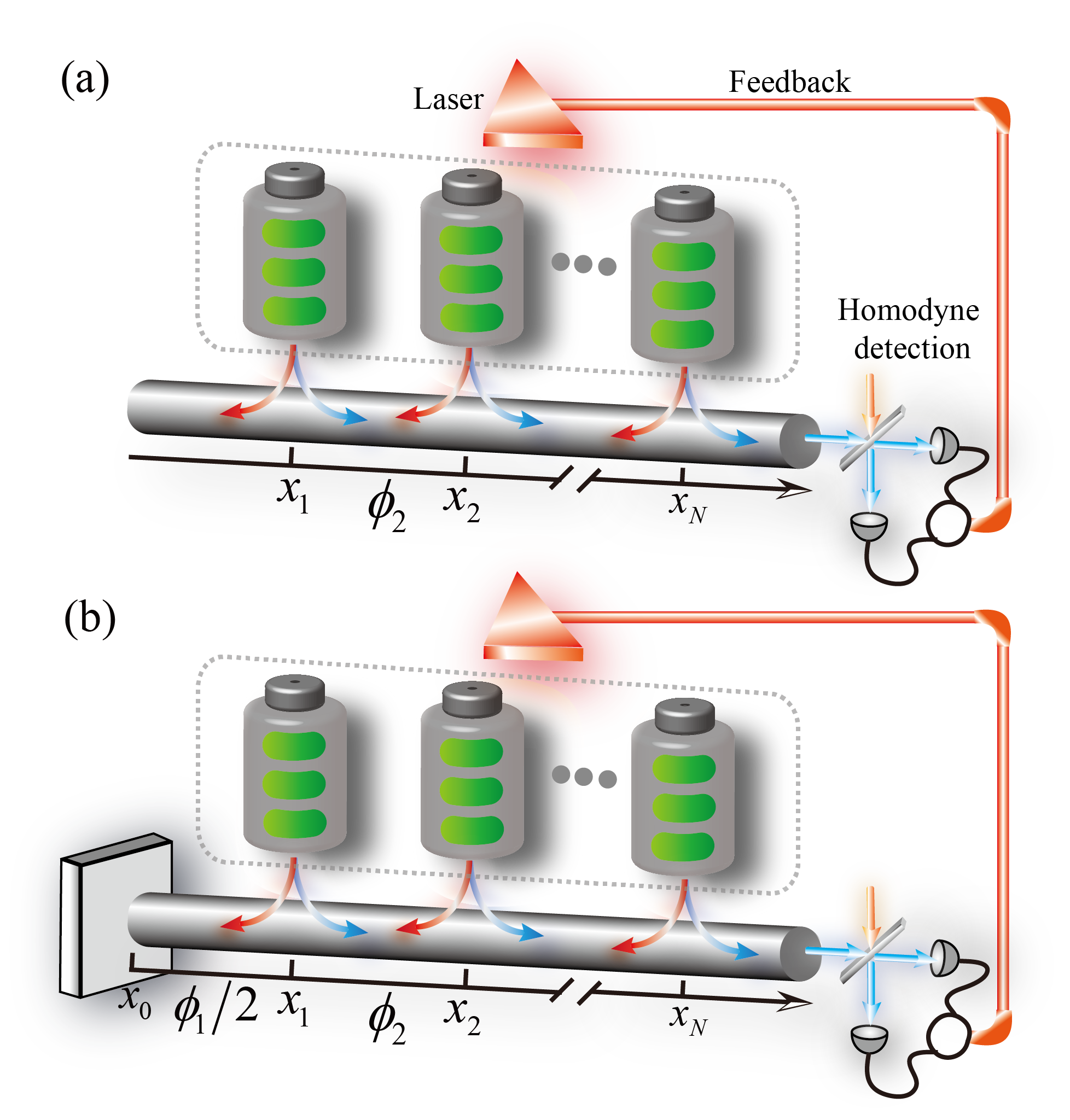}
\caption{Schematic of two different setups of the quantum battery (QB). (a) Setup I, a QB array (modeled as two-level atoms) is coupled to an open waveguide. (b) Setup II, a QB array is coupled to a semi-infinite waveguide, whose end lying at $x_{0}=0$ behaves as a perfect mirror. In both setups, an external laser persistently input energy into the QB array. The right-emitted photons are monitored via a homodyne detection, producing a corresponding photocurrent that is fed back to modulate the external laser field.}
\label{Fig1}
\end{figure}
%%%%%%%%%%%%%%%%%%%%%%%%%%%%%
\section{Driven-dissipative-charging QB model}
We consider that a QB array, modeled as an ensemble of two-level atoms with resonance frequency $\omega_{0}$, is coupled to either an open or a semi-infinite waveguide, as shown by setups I and II in Figs.~\ref{Fig1}(a) and~\ref{Fig1}(b), respectively. The single-atom spontaneous decay rates into the right- and left-propagating modes are $\gamma_{R}$ and $\gamma_{L}$, respectively, with a total decay rate $\gamma=\gamma_{R}+\gamma_{L}$. An external laser with driving amplitude $\Omega$ persistently input energy into the QB array to compete with the system dissipation. Photons emitted into the right-propagating mode are monitored via continuous homodyne detection of the phase quadrature $P(t)$ of the photocurrent. The measured photocurrent is then fed back into the system, which can instantaneously modify the driving amplitude of the laser. Within the Born--Markov approximation framework and using the SLH formalism~\cite{Gough091,Gough092,Zhang12,Combes17}, the quantum master equations describing the driven-dissipative dynamics of setups I and II are, respectively, given by~\cite{SM}
\begin{subequations}
\begin{align}
\dot{\rho}_{I}=\;&-i[H_{RL}+H_{f,I}+H_{d},\rho _{I}]+D[L_{R}-iF]\rho
_{I}+D[L_{L}]\rho _{I}\label{MEQS_NA_I}, \\  
\dot{\rho}_{II}=\;&-i[H_{RL}+H_{f,II}+H_{\phi_{1}}+H_{d},\rho _{II}]+D[e^{i\phi _{1}}S_{R}L_{L}+L_{R}-iF]\rho _{II}. \label{MEQS_NA_II}
\end{align}
\end{subequations}
The exchanging interaction mediated by the waveguide is given by $H_{RL}=\frac{\gamma _{R}}{2i}\sum_{j>l}e^{i\phi _{l,j}}\sigma_{j}^{+}\sigma _{l}^{-}+\frac{\gamma _{L}}{2i}\sum_{j<l}e^{i\phi_{l,j}}\sigma _{j}^{+}\sigma _{l}^{-}+\text{H.c.}$, where $\phi_{l,j}=k_{0}|x_{j}-x_{l}|$ being the propagating phase of photons between the positions $x_{l}$ and $x_{j}$, $k_{0}=\omega_{0}/\upsilon_{g}$ is the wave vector of the field at frequency $\omega_{0}$, and $\upsilon_{g}$ is the corresponding group velocity; $H_{f,I}=F^{\dagger }L_{R}/2+\text{H.c.}$ and $H_{f,II}=F^{\dagger }(e^{i\phi_{1}}S_{R}L_{L}+L_{R})/2+\text{H.c.}$ are the Hamiltonians induced by the MFB control in the case of setups I and II, respectively, where $F=i\sqrt{\gamma_{R}}g\sum_{i}\sigma_{i}^{x}$ denotes the feedback operator, with $g$ being the feedback strength; $D[o]\bullet=o\bullet o^{\dagger}-\{o^{\dagger}o,\bullet\}/2$ is the Lindblad superoperator, where the right and left collapse operators are, respectively, given by $L_{R}=\sqrt{\gamma _{R}}\sum_{j=1}e^{i\phi _{\Sigma _{R}}}\sigma_{j}^{-}$ and $L_{L}=\sqrt{\gamma _{L}}\sum_{j=1}e^{i\phi _{\Sigma _{L}}}\sigma_{j}^{-} $, with $\phi _{\Sigma _{R}}=\sum_{s=j+1}^{N}\phi _{s}$, and $\phi _{\Sigma _{L}}=\sum_{s=2}^{j}\phi _{s}$; $H_{\phi _{1}}=e^{i\phi_{1}}S_{R}L_{R}^{\dagger }L_{L}/2i+$H.c. is the Hamiltonian induced by the CFB control; $H_{d}=\Omega \sum_{j=1}^{N}\sigma _{j}^{x}$ is the driving Hamiltonian;  $S_{R}=S_{L}=e^{i\phi _{\Sigma }}$ are the scattering matrices for the right- and left-propagating modes, with $\phi _{\Sigma}=\sum_{s=2}^{N}\phi _{s}$.

To characterize the performance of the driven-dissipative QB in setups I and II, we introduce the stored energy at time $t$, defined as $\mathcal{E}(t)=\text{Tr}[H_{B}\rho _{B}(t)]$. Here, $H_{B}$ and $\rho(t)$ are, respectively, the bare energy and density matrix of the atoms. This characterization is justified under the typical condition where $\omega_{0}\gg\gamma,\Omega$~\cite{Zoller20,Kannan23,Shah24}. The second law of thermodynamics shows that not all of $\mathcal{E}(t)$ is available for work extraction. The maximum work that can be extracted from the state $\rho(t)$ of  the QB is called ergotropy, defined as $\mathcal{W}(t)=\text{Tr}[\rho(t)H_{B}]-\text{Tr}[\sigma (t)H_{B}]$~\cite{Allahverdyan04,Francica20}, where $\sigma(t)=\sum_{n}r_{n}(t)|\epsilon_{n}\rangle\langle\epsilon_{n}|$ is the passive state. Here, $r_{n}(t)$ are the eigenvalues of $\rho(t)$ sorted in decreasing order, $r_{1}\geq r_{2}\geq \cdot\cdot\cdot\geq r_{N}$, and $|\epsilon _{n}\rangle $ are the eigenstates of $H_{B}$ with the corresponding eigenvalues $\epsilon_{n}$ sorted in an increasing order, $\epsilon_{1}\leq \epsilon_{2}\leq \cdot\cdot\cdot\leq \epsilon_{N}$.

\section{Feedback-enhanced single-atom QB}
In order to elucidate the charging mechanism in the presence of feedback control, we first focus on the single-atom driven-dissipative QB. According to Eq.~(\ref{MEQS_NA_I}), the dynamics of the QB in setup I is given by the Lindblad master equation
\begin{equation}
\label{MEQS_1atom}
\dot{\rho}_{I}=i[\rho _{I},\Omega \sigma ^{x}]+\sum_{\mu ,\nu=+,-}\Gamma _{\mu \nu }^{(I)}D_{\sigma ^{\mu },\sigma ^{\nu }}\rho _{I},
\end{equation}
where the feedback-strength-dependent decay rates are given by $\Gamma _{-+}^{(I)}=\gamma_{R}(g+1)^{2}+\gamma _{L}$, $\Gamma _{+-}^{(I)}=\gamma _{R}g^{2}$, and $\Gamma _{++}^{(I)}=\Gamma _{--}^{(I)}=\gamma _{R}g(g+1)$, and $D_{\sigma^{\mu},\sigma^{\nu}}\bullet=\sigma^{\mu}\bullet\sigma^{\nu}-\{\sigma^{\nu}\sigma^{\mu},\bullet\}/2$. It is interesting to see that the master equation~(\ref{MEQS_1atom}) has the same form as that of a driving two-level atom in a squeezed vacuum reservoir~\cite{Gardiner86,Walls87,Siddiqi16,Yin25}. By adjusting the feedback strength $g$, we can enhance or suppress the decay process of the QB and hence achieve different stable charging.

From the decay rates in Eq.~(\ref{MEQS_1atom}), one can find that the components $\gamma_{L}$ in $\Gamma_{-+}^{(I)}$ is independent of $g$. In our scheme, only the right-propagating photons are measured by the homodyne detector. As a result, the jump operator associated with left-propagating modes remains unchanged. To address this issue, we further investigate the charging performance in setup II, where the waveguide end behaves as a perfect mirror. The photons emitted to the left-propagating modes will be reflected back to the atoms, and hence the dynamics of the atoms can also be controlled by the CFB control~\cite{Tufarelli13,Wilson15,Ciccarello19}. The master equation for the single-atom QB in setup II is obtained as
\begin{equation}
\dot{\rho}_{II}=i[\rho _{II},\Omega \sigma ^{x}+\omega \sigma ^{+}\sigma^{-}]+\sum_{\mu ,\nu =+,-}\Gamma _{\mu \nu }^{(II)}D_{\sigma ^{\mu},\sigma ^{\nu }}\rho _{II},
\end{equation}
with the frequency shift $\omega =\sqrt{\gamma _{R}\gamma_{L}}\sin \phi _{1}(g+1)$ and decay rates $\Gamma _{-+}^{(II)}=\gamma
+\gamma _{R}g(g+2)+2\sqrt{\gamma _{R}\gamma _{L}}(g+1)\cos \phi _{1}$, $\Gamma _{+-}^{(II)}=\gamma _{R}g^{2}$, and $\Gamma _{++}^{(II)}=(\Gamma
_{--}^{(II)})^{\ast }=\gamma _{R}g(g+1)+\sqrt{\gamma _{R}\gamma _{L}}ge^{-i\phi _{1}}$. Owing to the CFB control, the decay rate associated with the left-propagating mode can also be adjusted by the feedback strength. The CFB control is engineered by the propagating phase $\phi_{1}$, which can enhance or suppress the decay of the QB dynamics. In the absence of the MFB control, if the phase is fixed at $\phi_{1}=\pi$, the two-level atom is completely decoupled from the waveguide~\cite{Tufarelli13,Wilson15,Ciccarello19}, and hence the energy stored in the QB can exhibit a periodical oscillation, with a period determined by the driving frequency of the laser. In the driven-dissipative regime, we fix $\phi_{1}=2\pi$ and consider the achiral (bidirectional) coupling case ($\gamma_{R}=\gamma_{L}=\gamma/2$). Under these conditions, we can investigate the combined effects of the driving, dissipation, and feedback controls, while the chiral-coupling case is discussed in the Supplementary Information~\cite{SM}.

%%%%%%%%%%%%%%%%%%%%%%%%%%%%%
\begin{figure}[t]
\center\includegraphics[width=0.49\textwidth]{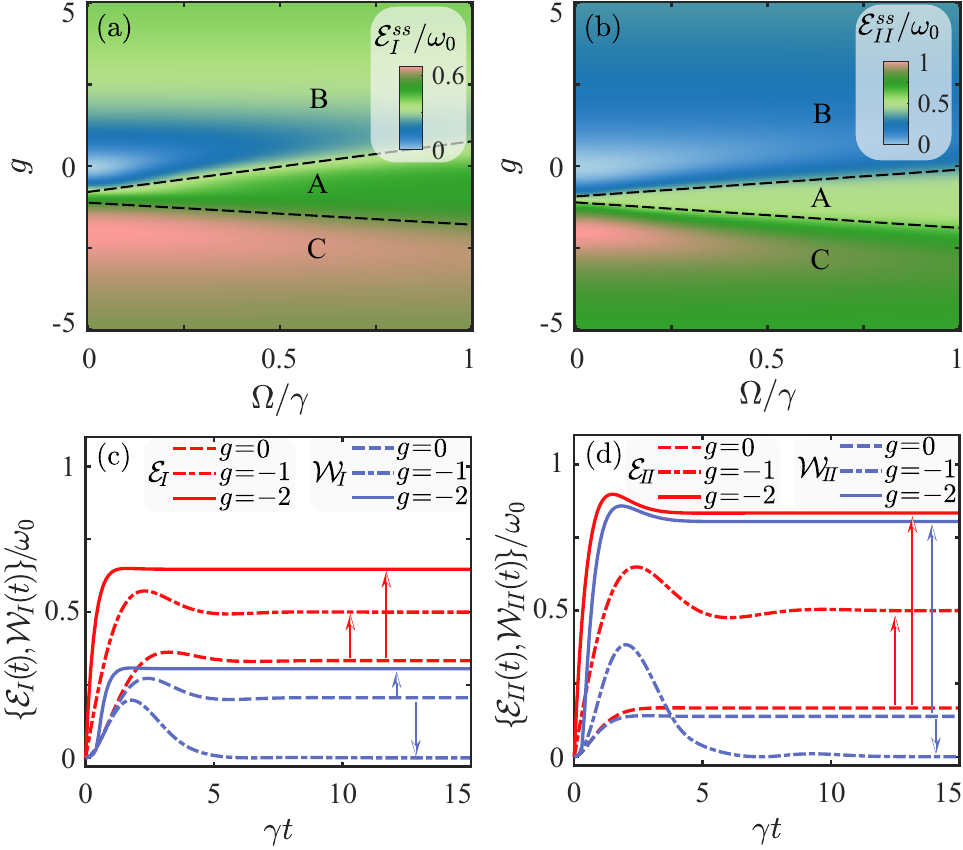}
\caption{Steady-state energy $\mathcal{E}^{ss}_{I}$ (a) and $\mathcal{E}^{ss}_{II}$ (b) versus the driving amplitude $\Omega$ and feedback strength $g$. Time evolutions of the energy and ergotropy for setups I (c) and II (d), comparing the cases with $g=0$ (without MFB) and $g=-1$, $-2$ (with MFB). In panels (c) and (d), the driving amplitude is fixed at $\Omega/\gamma=0.5$. Other parameters used are $\Delta\gamma=0$ and $\phi_{1}=2\pi$.}
\label{Fig2}
\end{figure}
%%%%%%%%%%%%%%%%%%%%%%%%%%%%%

From the dynamics of the density matrix elements, the QB in the long-time limit evolves into the steady state
\begin{equation}
\rho_{i}^{ss}=\frac{1}{\eta _{i}}\left(
\begin{array}{cc}
\varsigma_{i} & -4i(g+1)\gamma \Omega  \\
4i(g+1)\gamma \Omega  & \eta _{i}-\varsigma _{i}
\end{array}
\right),
\end{equation}
where $i=\text{I}$ and $\text{II}$ are used to label the two different QB setups. The  two eigenvalues of $\rho _{i}^{ss}$ are given by $r_{\pm}=1/2\pm\chi _{i}/\eta _{i}$. The specific forms of the parameters $\eta _{i}$, $\varsigma_{i}$, and $\chi _{i}$ are presented in the SM ~\cite{SM}, which strongly depend on the driving amplitude $\Omega$ and the MFB control strength $g$.

The steady-state energy-storage performance of the QB in setups I and II is evaluated as a function of the driving amplitude $\Omega$ and MFB control strength $g$ in Figs.~\ref{Fig2}(a) and~\ref{Fig2}(b),  where we see that the feedback strength $g$ approximately partitions the energy landscape into three distinct regions, labeled A, B, and C, respectively. The steady-state energies in regions B and C are significantly influenced by both $\Omega$ and $g$, with region C supporting the highest steady-state energy. In contrast, the energy in region A is largely insensitive to these parameters. The comparison of $\mathcal{E}_{I}^{ss}$ and $\mathcal{E}_{II}^{ss}$ further indicates that the QB in setup II exhibits superior energy-storage performance in the driven-dissipative regime, which achieves a larger steady-state energy in a fairly wide parameter regime. Furthermore, under the weak driving case $\Omega/\gamma\ll1$, the QB in setup II can approach the full-charge limit.

Figs.~\ref{Fig2}(c) and~\ref{Fig2}(d) compare the time evolutions of the energy and ergotropy for these two setups. When $g=-2$, the MFB control significantly boosts the charging process and enhances the maximal ergotropy in both cases. For setup II, the evolution of $\mathcal{W}_{II}(t)$ closely follows that of $\mathcal{E}_{II}(t)$, indicating high extractable work of the QB. For setup I, however, the steady-state ergotropy is much smaller than the stored energy due to losses into the left-propagating waveguide mode. A critical finding is that at $g=-1$, the steady-state density matrix becomes fully mixed, with vanishing coherence (off-diagonal elements). Consequently, the steady-state energy saturates at $1/2$, and no work can be extracted from the QB. These results show that by appropriately tuning the feedback strength $g$ and driving amplitude $\Omega$, the charging performance of a single-atom QB can be effectively enhanced.

\section{Many-body phase-controlled QB array}
After establishing the role of the feedback control in enhancing the single-atom QB, we now extend our analysis to the many-body charging behavior of $N$ two-level atoms serving as a QB array. In this case, the MFB control can be used to engineer rich dynamical phases due to the multiple-atom collective effect~\cite{Buonaiuto21,Ivanov20}. By positioning the atoms to satisfy $\phi_{i}=2m\pi$ ($i=1, 2,\cdots,N$ and $m\in\mathbb{Z}$), the quantum master equations in Eqs.~(\ref{MEQS_NA_I}) and~(\ref{MEQS_NA_II}) are reduced to the unified form
\begin{equation}
\label{MATs_MQES_I_II}
\dot{\rho}=i\left[\rho ,2\Omega J_{x}-\frac{n\gamma g}{2}\{J_{x},J_{y}\}\right]+\sum_{\mu ,\nu =+,-}\Gamma _{\mu \nu }D_{J_{\mu },J_{\nu }}\rho ,
\end{equation}
where $J_{\alpha=x,y,z}=\sum_{j=1}^{N}\sigma_{j}^{\alpha}/2$ and $J_{\pm}=J_{x}\pm iJ_{y}$ are the collective operators. The index $n=1$ and $2$ correspond to setups I and II, respectively, with the decay rates $\Gamma _{-+}=\gamma[g^{2}+2n(g+1)]/2$, $\Gamma _{+-}=g^{2}/2$, and $\Gamma _{++}=\Gamma _{--}=g(g+n)/2$. Therefore, the constructive interference of propagating photons in the waveguide yields a collective spin mode of length $N/2$~\cite{Agarwal90,Song17,Bai21}.

%%%%%%%%%%%%%%%%%%%%%%%%%%%%%
\begin{figure}[t!]
\center\includegraphics[width=0.49\textwidth]{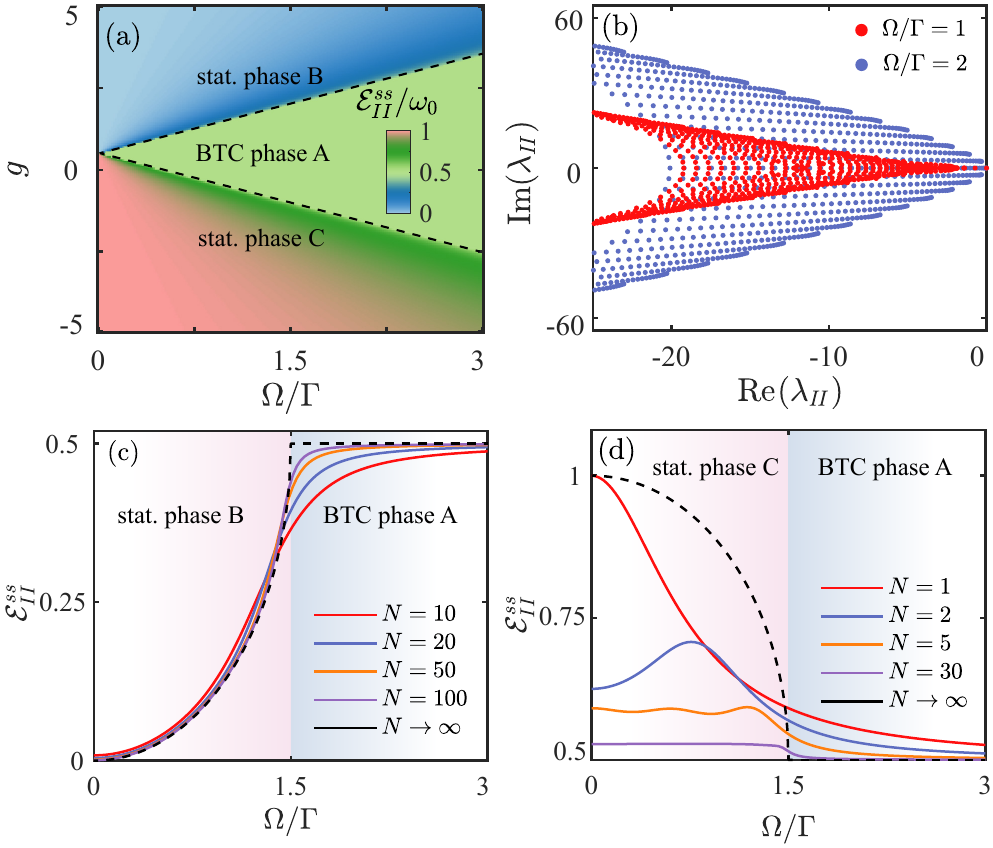}
\caption{(a) Steady-state energy $\mathcal{E}_{II}^{ss}$ versus $\Omega$ and $g$. (b) Eigenvalues $\lambda_{II}$ of the Liouvillian $\mathcal{L}$ are shown in the boundary time-crystal phase A ($\Omega/\Gamma=2$, blue dots) and in the stationary phase B ($\Omega/\Gamma=1$, red dots) with $g=1$. (c) and (d) Steady-state energy $\mathcal{E}_{II}^{ss}$ as a function of $\Omega$ at $g=1$ and $g=-2$, respectively. Dashed lines are the mean-field results for $N\rightarrow\infty$ and solid lines are obtained by finding the steady state of Eq.~(\ref{MATs_MQES_I_II}) for setup II with various atom numbers $N$.}
\label{Fig3}
\end{figure}
%%%%%%%%%%%%%%%%%%%%%%%%%%%%%

In the QB array, the stored energy is defined as $\mathcal{E}/N\omega_{0}=(\langle J_{z}\rangle/N)+1/2$, with the expectation value defined as $\langle\bullet\rangle=\text{Tr}[\bullet\rho]$. We consider the thermodynamics limit $N\rightarrow\infty$ and introduce the expectation values for the magnetization vector components  $m_{\alpha}=\langle J_{\alpha}\rangle/N$~\cite{Fazio18}. When $N\rightarrow\infty$, the dynamics of the expectation values is exactly governed by the mean-field equations~\cite{SM}: $\dot{m}_{x} =n\Gamma m_{z}m_{x}$, $\dot{m}_{y} =-2\Omega m_{z}+n\Gamma\xi m_{y}m_{z}$, and $\dot{m}_{z}=2\Omega m_{y}-n\Gamma m_{x}^{2}-n\Gamma\xi m_{y}^{2}$, with $\xi=2g+1$. To ensure a well-defined thermodynamic limit, we introduce the rescaled decay rate  $\Gamma=N\gamma$ for the $N/2$ spin. The mean-field equations conserve the length $j$ of the average magnetization vector $\vec{m}=[m_{x}(t),m_{y}(t),m_{z}(t)]$. In addition, the system exists another conserved quantity $\mathcal{M}\equiv \Gamma m_{x}^{\kappa}/(\Gamma \kappa m_{y}-\Omega)$~\cite{SM}, which thus constrains the magnetization dynamics to a specific trajectory. For the many-body charging process, we consider that the $N$ atoms are initially in their ground state, with the initial condition $m_{x}(0)=m_{y}(0)=0$ and $m_{z}(0)=-1/2$. Therefore, we have $j=1/4$ and $\mathcal{M}=0$. It is found that there exists two stationary phases and a BTC phase in both setups I and II. When the driving amplitude satisfies $\Omega>\Omega_{\text{cri}}=n\Gamma|\xi|/4$, the energy stored in the QB array will exhibit persistent oscillations even when the atoms decay to the fields in the waveguide. In contrast, the time evolution of the stored energy in the QB array approaches two stationary values $\mathcal{E}^{ss}=-\text{sign}(\xi)\sqrt{1/4-4\Omega^{2}/n^{2}\Gamma ^{2}\xi ^{2}}+1/2$~\cite{SM}, whose sign depends on the feedback strength. For $\xi \geq 0$ (i.e., $g \geq -1/2$), we find that $\mathcal{E}^{ss}\leq 1/2$ holds for both setups I and II. This behavior is consistent with the steady-state charging energy achieved via collective effect in the driven-dissipative regime without feedback control~\cite{Chang21}. While for $\xi < 0$ (i.e., $g < -1/2$), the relation $\mathcal{E}^{ss}> 1/2$ holds for both setups. The stability of these two stationary solutions can be checked by examining the eigenvalues of the Jacobian matrix $J$~\cite{Mattes23}. We see that the critical driving amplitude $\Omega_{\text{cri}}$ and the steady-state stored ennergy $\mathcal{E}^{ss}$ in setups I and II have the similar form in the thermodynamic limit. Therefore, we focus on the charging behavior of the QB array in setup II. A comparison of the charging performance for these two setups in finite system size is presented in the Supplementary Information~\cite{SM}.

Figure~\ref{Fig3}(a) illustrates that the steady-state energy $\mathcal{E}_{II}^{ss}$ is well divided into three distinct dynamical regions by the critical driving amplitude $\Omega_{\text{cri}}$ (as shown by the black dashed lines):  BTC phase A, and stationary phases B and C. As we will show, the QB array exhibits persistent oscillatory dynamics in the phase A, while it relaxes to two distinct steady-state in phases B and C, with the stationary phase C yielding a higher steady-state charging energy.

The emergence of the different dynamical phases can be seen from the properties of the eigenvalues $\lambda_{II}$ of the Liouvillian superoperator $\mathcal{L}$~\cite{Fazio18}. The quantum master equation~(\ref{MATs_MQES_I_II}) can be written compactly as $\dot{\rho}=\mathcal{L}\rho$, with the formal solution $\rho(t)=e^{\mathcal{L}t}\rho(0)=\sum_{s}c_{s}e^{\lambda_{s}t}\rho_{s}$, where $\lambda_{s}$ denotes the $s$th eigenvalue of $\mathcal{L}$, $\rho_{s}$ is the corresponding eigenmode, and the coefficient $c_{s}$ is determined by the initial state. As shown in Fig.~\ref{Fig3}(b), the Liouvillian spectrum in the stationary phase (red dots) is gapless. The eigenvalue with the largest real part possesses a nonzero imaginary component, causing the QB dynamics to approach a stationary value after transient oscillations. In contrast, the spectrum in the BTC phase (blue dots) is gapped, and the leading eigenvalue has a vanishing imaginary part, giving rise to sustained coherent oscillations.

A comparison between the exact numerical results of the master equation~(\ref{MATs_MQES_I_II}) and the mean-field predictions is studied in Figs.~\ref{Fig3}(c) and~\ref{Fig3}(d). As shown in Fig.~\ref{Fig3}(c), the mean-field results converge toward the exact numerics as the atom number $N$ increases, and the steady-state energy $\mathcal{E}_{II}^{ss}$ approaches $0.5$ with increasing $\Omega$. However, Fig.~\ref{Fig3}(d) reveals that while $\mathcal{E}_{II}^{ss}$ converges toward the mean-field prediction in the BTC phase, its behavior in stationary phase C appears to be different. In particular, a significant deviation from the mean-field results emerges for system sizes achievable in steady-state calculations.

%%%%%%%%%%%%%%%%%%%%%%%%%%%%%
\begin{figure}[t]
\center\includegraphics[width=0.49\textwidth]{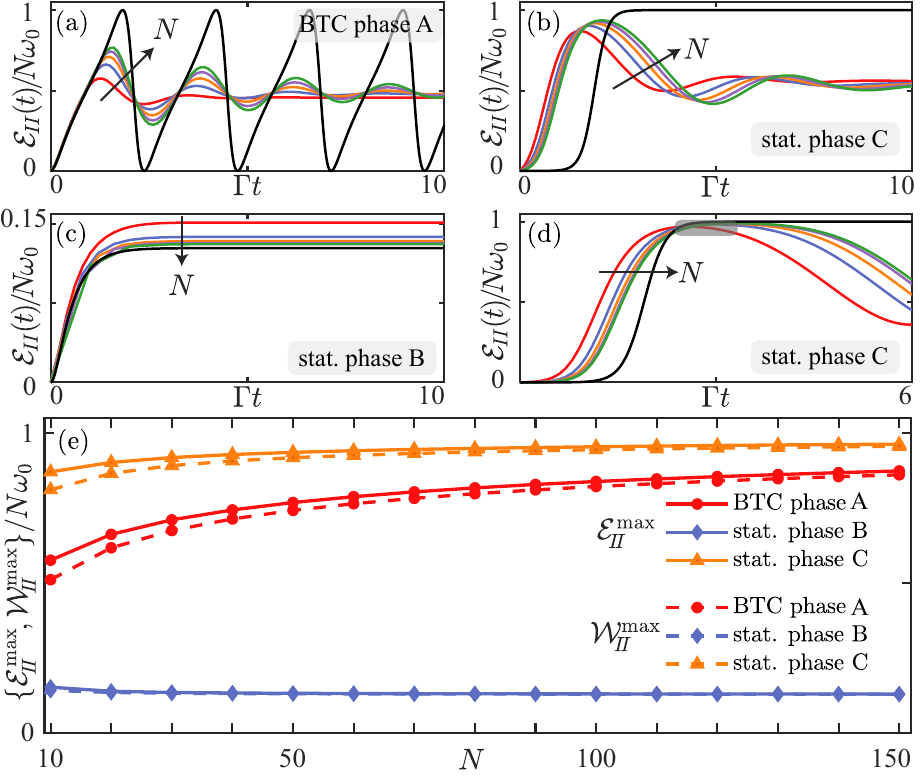}
\caption{(a) and (c) Stored energy $\mathcal{E}_{II}(t)$ as a function of time in the BTC phase A and stationary phase B for $\Omega/\Gamma=1$ and $\Omega/\Gamma=2$ with $g=1$. (b) and (d) Stored energy $\mathcal{E}_{II}(t)$ as a function of time in the stationary phase C for $\Omega/\Gamma=0.01$ and $g=-2$. Panels (a), (b), and (c) show $N=10$, $20$, $30$, $40$, $50$. Panel (d) shows instead more extensive comparison for $N=200$, $600$, $1000$, $1400$, $1800$. (e) Maximal energy $\mathcal{E}_{II}^{\text{max}}$ and ergotropy $\mathcal{W}_{II}^{\text{max}}$ versus $N$ in different phases. }
\label{Fig4}
\end{figure}
%%%%%%%%%%%%%%%%%%%%%%%%%%%%%

To investigate whether this deviation persists in larger system size, we examine the finite time evolution of the energy in Figs.~\ref{Fig4}(a)-\ref{Fig4}(d), which allows us to access larger atom numbers $N$ within feasible computation time. In the BTC phase, $\mathcal{E}_{II}(t)$ exhibits damped oscillations until it approaches the steady-state value $0.5$ for finite atom numbers $N$. As $N$ increases, $\mathcal{E}_{II}(t)$ exhibits persistent oscillations. In the stationary phases B, $\mathcal{E}_{II}(t)$ fast approaches a small steady-state value, which converges to the mean-field results with the increase of $N$. In the stationary phase C, we observe that for small system size, $\mathcal{E}_{II}(t)$ has a larger derivation from the mean-field results as time increases. 
Nevertheless, this does not invalidate the mean-field method, which is exact only in the thermodynamic limit, despite the fact that all experimental systems are inherently finite~\cite{Carollo24}. For a comparison, we show $\mathcal{E}_{II}(t)$ in the stationary phase C for larger $N$ in Fig.~\ref{Fig4}(d) over a shorter time window. This shows how the exact numerics indeed approach the mean-field prediction for finite-$N$ simulations. Owing to the existence of the MFB control, the stored energy in the QB array is close to full charge in the case of $\Omega/\Gamma\ll1$, marked by the shade area in Fig.~\ref{Fig4}(d). Figure~\ref{Fig4}(e) shows the maximal $\mathcal{E}_{II}^{\text{max}}$ and ergotropy $\mathcal{W}_{II}^{\text{max}}$ versus the atom numbers $N$ in the three different phases. For a fixed value of $N$, both $\mathcal{E}_{II}^{\text{max}}$ and $\mathcal{W}_{II}^{\text{max}}$ are the highest in the stationary phase C. As $N$ increases, these quantities remain nearly constant but small in the phase B. In contrast, in the stationary phase C and BTC phase, $\mathcal{E}_{II}^{\text{max}}$ and $\mathcal{W}_{II}^{\text{max}}$ gradually approach unity with increasing $N$. In all phases, $\mathcal{W}_{II}^{\text{max}}$ converges toward $\mathcal{E}_{II}^{\text{max}}$ as $N$ grows.

\section{Conclusions and outlook}
In conclusion, we have explored the charging performance of the driven-dissipation QB with MFB control in two different setups based on waveguide-QED systems. In the single-atom-QB scenario, the steady-state energy and ergotropy of the QB can be remarkably enhanced by the MFB control. Extending to the multiple-atom case, we showed that the MFB control and the laser driving cause the steady-state energy charging to exhibit three different dynamical phases. These rich dynamical phases provide us a valid method to control the charging performance of the QB on demand. Our results advance the realization of QBs and provide a guideline for manipulating the charging process in open quantum systems. Future work could investigate physically richer settings for QB charging in the driven-dissipative regime, such as exploring the role of time delays between coupling points of emitters~\cite{Pichler16,Vodenkova24,Windt25,Pascual25}, using the rich interference effects in giant atoms~\cite{Kockum21}, and extending the scheme to multi-level atoms~\cite{Iversen21,Poddubny24}.

\begin{acknowledgements}
We thank F. Carollo, X.-Z. Ge, and Y.-H. Zhou for helpful discussions. We also acknowledge the use of the open source Python numerical packages QuTiP~\cite{Johansson12,Johansson13}. This work is partially financially supported by Innovation Program for Quantum Science and Technology 2023ZD0300600, Guangdong Provincial Quantum Science Strategic Initiative (No.~GDZX2200001), Hong Kong Research Grant Council (RGC) under Grant No.~15213924, National Natural Science Foundation of China under Grants No.~62173288.
\end{acknowledgements}

\end{document}